\documentclass[aip,pof,preprint,tightenlines]{revtex4-1}

\usepackage{graphicx}
\usepackage{amssymb}
\usepackage{amsmath}
\usepackage{graphicx}
\usepackage{MnSymbol}

\begin{document}
\title{ Reshaping and capturing Leidenfrost drops with a magnet}

\author{ Keyvan Piroird}
\author{Baptiste Darbois Texier}
\author{Christophe Clanet}
\author{David Qu\'er\'e}
\affiliation{PMMH, UMR 7636 du CNRS, ESPCI, 75005 Paris, France}
\affiliation{LadHyX, UMR 7646 du CNRS, \'Ecole Polytechnique, 91128 Palaiseau, France}

\begin{abstract}
Liquid oxygen, which is paramagnetic, also undergoes Leidenfrost effect at room temperature. In this article, we first study the deformation of oxygen drops in a magnetic field  and show that it can be described via an effective capillary length,  which includes the magnetic force. In a second part, we  describe how these ultra-mobile drops passing above a magnet significantly slow down and can even be trapped. The critical velocity below which a drop is captured is determined from the deformation induced by the field. 
\end{abstract}

 \maketitle 

\section{Introduction}
Liquid oxygen is known for its paramagnetic properties since the  pioneering work of James Dewar, who first noticed that it is attracted by the poles of an electromagnet \cite{Dewar1927}. Moreover, because of its low boiling point ($T=-183^{\circ}$C at atmospheric pressure), an oxygen drop on a substrate at room temperature rapidly evaporates, forming a cushion of vapour on which it levitates, a phenomenon known as the Leidenfrost effect, that was reported for the first time in 1756  \cite{Leidenfrost1756, Gottfried1966} and that continues to inspire research nowadays \cite{Thoroddsen2012,Quéré2013}.  A Leidenfrost drop achieves a perfect non-wetting situation, where there is no contact between the liquid and its solid support. As a consequence, drops adopt a very rounded shape and they are extremely mobile. Despite these remarkable properties, liquid oxygen has been much less described than other magnetic fluids such as ferrofluids (colloidal suspension of ferromagnetic nanoparticles), which are liquid at room temperature and have a much higher magnetic susceptibility (roughly 100 times higher than liquid oxygen). Liquid oxygen has nonetheless been studied in the framework of magnetic compensation of gravity \cite{Catherall2003,Nikolayev2009}, surface instabilities \cite{Takeda1991,Catherall2003Instabilities}, and pumping with magnetic field \cite{Youngquist2003,Boulware20102}.  Density and surface tension of liquid oxygen at the boiling point ($T=-183^{\circ}$C) are $\rho=1140~\mathrm{kg/m^3}$ and $\gamma=13$ mN/m. Takeda \emph{et al.} \cite{Takeda1991ST} showed that  $\gamma$ remains constant under a uniform magnetic field up to $5\;$T. In this article, we study how the shape and mobility of oxygen drops are influenced by a magnet, which allows us to control them in a non-intrusive way. In a first part, we study how the shape of an oxygen drop is modified in a magnetic field, and we relate in a second part this deformation to the slowing down and capture of mobile drops.

\section{Static shape of oxygen drops}
\begin{figure}[h!]
	\centering
\includegraphics[width=16cm]{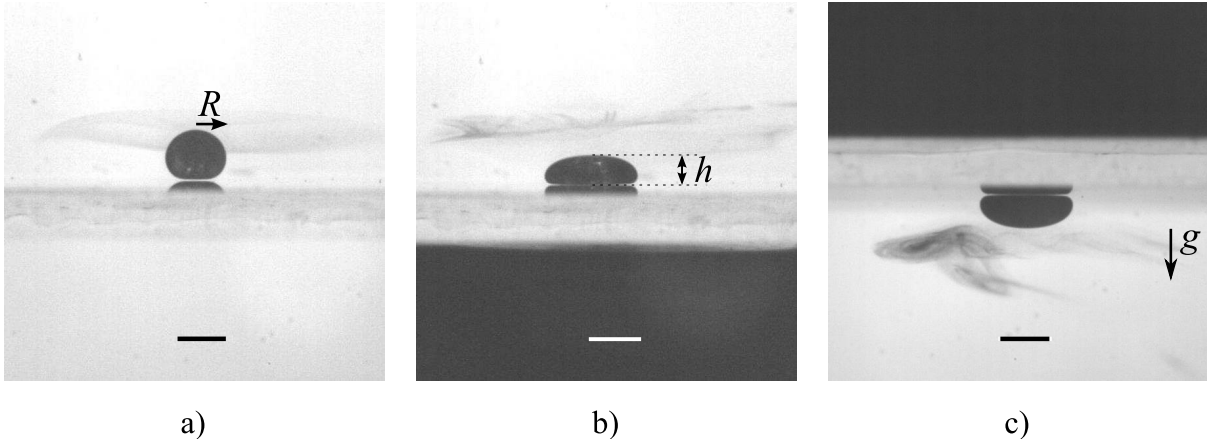}
\caption{ Side views of liquid oxygen drops on a glass plate. a) Liquid oxygen drop of equatorial radius $R=0.7\;$mm on a glass plate at room temperature. b) The same drop in the presence of a magnet (in black, in the picture) $1\;$mm below it. The drop is deformed and is no longer spherical. c) Same set-up as before turned upside down. The drop holds against gravity and its shape is almost the same as is b). The solid line represents 1 mm in each picture.}
\label{fig:StaticDeformation}
\end{figure}

Liquid oxygen is obtained by distillation of air using liquid nitrogen, which boils at $-196^{\circ}$C. A copper sheet of millimetric thickness is folded and welded to obtain a cone of about $10\;$cm  height and width. It is then filled with liquid nitrogen: the cone temperature quickly reaches $-196^{\circ}$C, that is, $13^{\circ}$C below the boiling point of oxygen present in the air, which therefore liquefies on the external surface of the cone. Liquid oxygen drains along the copper and drips at the tip, where it is collected and directly used. Along with oxygen, other components of air presenting a condensed phase at this temperature might be present. Argon, which liquefies at $-186^{\circ}$C and solidifies at $-189^{\circ}$C, should represent $5\%$ in the liquid obtained at the surface of the cone. Water should be present as solid in typically the same proportion, and ice crystals are indeed observed, making the drop milky, and thus uniformly dark when backlighted (Fig.~\ref{fig:StaticDeformation}). Carbon dioxide is also solid at this temperature, but it only represents a proportion on the order of 1 \textperthousand.

Without magnetic field, the shape of a levitating oxygen drop is dictated by a balance between gravity and capillarity. As seen in Figure \ref{fig:StaticDeformation}a, a drop of radius $R$ smaller than the capillary length $a=\sqrt{\gamma/\rho g}= 1.1\;$mm, is almost spherical except for a small region at the bottom, of typical size $R^2/a~$ \cite{Maha1999}. Thanks to surface reflection, we observe the presence of the vapour film, of typical thickness $\delta\sim50\;\mu$m, between the drop and its support. A small cloud is visible around the drop resulting from the condensation of water vapour present in air. Large drops ($R>a$) are deformed by gravity and look like puddles (not shown here), whose thickness $h$ is fixed by a balance between gravity and capillarity: in the limit where $h\ll R$, we can approximate the shape of this gravity-dominated puddle by a cylinder of volume $\Omega=\pi R^2 h$, whose total energy $E\simeq2\gamma \Omega/h + \rho g \Omega h/2$ is minimal for $h=2a$.

A first experiment consists of approaching a magnet below an oxygen drop such as the one in Figure \ref{fig:StaticDeformation}a. We use a cylindrical neodymium magnet, $3\;$cm in diameter and $1\;$cm thick, generating a magnetic field $B=0.5\;$T at the surface of the magnet (measured with a Hall effect teslameter) and decreasing on a length-scale similar to the size of the magnet. As seen in Figure \ref{fig:StaticDeformation}b, the drop is flattened and looks like a puddle, as if the presence of the magnet had modified the capillary length of oxygen. Figure \ref{fig:StaticDeformation}c shows the same set-up turned upside down, holding the drop against gravity. The shape of the drop is almost the same as before, suggesting that magnetic effects dominate gravity. The magnet exerts a force per unit volume \cite{Rosensweig}:
\begin{equation}
\boldsymbol{f_{m}}=\frac{\chi}{2 \mu_o} \boldsymbol{\nabla} \left(B^2 \right)
\label{eq:force}
\end{equation}
\noindent
\begin{figure}[b!]
	\centering
		\includegraphics[height=4.6cm]{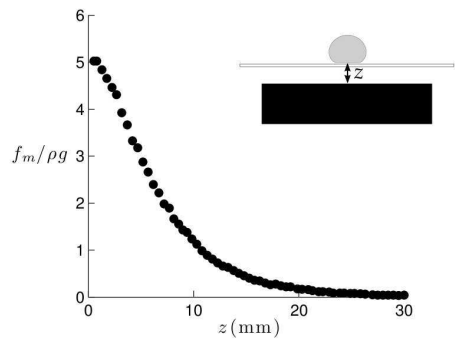}
\caption{\footnotesize Modulus of the magnetic force per unit volume $f_{m}$ (deduced from a measurement of the magnetic field), normalized by the volumic weight of the drop $\rho g$, as a function of $z$, the distance between the magnet and the bottom of the drop as sketched in the insert.}
\label{fig:StaticForce}
\end{figure}
where $\chi$ is the magnetic susceptibility of liquid oxygen ($\chi=0.0035$ at $-183^{\circ}$C), $\mu_o=4\pi\cdot10^{-7}~\mathrm{H.m^{-1}}$ the magnetic permeability of vacuum and $B$ the modulus of the magnetic field. The quadratic dependence with the magnetic field comes from the interaction between the imposed field $B$ and the induced magnetization of liquid oxygen, proportional to $B$. This conservative force derives from a magnetic energy per unit volume:
 \begin{equation}
 E_m=-\chi B^2 / 2 \mu_o
\label{eq:Em}
 \end{equation}
  \noindent
which is always negative for liquid oxygen and proportional to the square of the magnetic field, meaning that the drop is equally attracted by both poles of a magnet. In our experiment, the magnet is ten times larger than the drop. The magnetic field is therefore homogeneous in the horizontal plane and only depends on the vertical coordinate. We call $z$ the distance between the magnet and the bottom of the drop. The value of the magnetic force deduced from the measurement of $B(z)$ above the magnet is reported in Figure~\ref{fig:StaticForce}.

Far from the magnet ($z>20\;$mm), the magnetic force is negligible compared to the volumic weight $\rho g$. For $z\approx10\;$mm, these two forces are on the same order, and the ratio $f_m /\rho g$ can go up to five at a few millimeters from the magnet. This ratio can even be higher for smaller magnets, for which the magnetic field gradient is stronger. Since the magnetic force acts in the same direction as gravity, we define a modified capilllary length $a^*$ to account for the change of shape observed in Figure \ref{fig:StaticDeformation}.
\begin{equation}
a^*(z)=\sqrt{\frac{\gamma}{\rho g + f_m(z)}} = \frac{a}{\sqrt{1+\frac{f_m(z)}{\rho g}}}
\label{eq:astar}
\end{equation}

\begin{figure}[b!]
	\centering
		   \includegraphics[height=5.5cm]{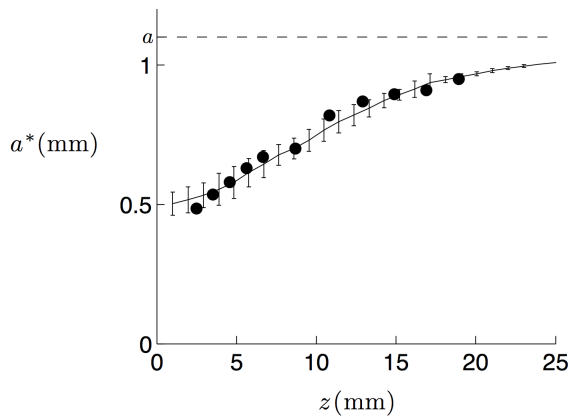}
	\caption{\footnotesize Magneto-capillary length $a^*$ as a function of $z$. The solid line is equation \eqref{eq:astar}, where $f_m$ is deduced from the measurement of the magnetic field. Error bars on this line arise from the discrete number of points where $B$ is measured, and are therefore larger where the field rapidly varies. The black dots show  half the thickness ($a^*\approx h/2$) of oxygen puddles. The dashed line represents the capillary length $a=1.1\;$mm without field.}
	\label{fig:OxygeneCalefieLongueurMagnetocapillaire}
\end{figure}
This modified capillary length reduces to the standard capillary length $a$ when $z$ is large, and it decreases as we get closer to the magnet. This explains the change of shape observed in Figure \ref{fig:StaticDeformation}: placed at $z=1\;$mm above the magnet where $a^*\simeq 0.5\;$mm, a drop of radius $R=0.7\;$mm is larger than $a^*$ and thus cannot remain spherical. If $R$ is initially larger than $a$, the drop is flat and the presence of a magnetic field changes the thickness of the puddle, which is twice the capillary length. Measuring this thickness as a function of $z$ gives us a direct measurement of $a^*(z)$, that we can compare to the value obtained from equation \eqref{eq:astar}, where $f_m$ is deduced from Figure \ref{fig:StaticForce}. These results, shown in Figure \ref{fig:OxygeneCalefieLongueurMagnetocapillaire}, are in good agreement with each other. The capillary length can therefore be varied continuously by a factor two without changing the surface tension nor the density of the liquid.

\begin{figure}[h!]
 		\centering
  		\includegraphics[height=4cm]{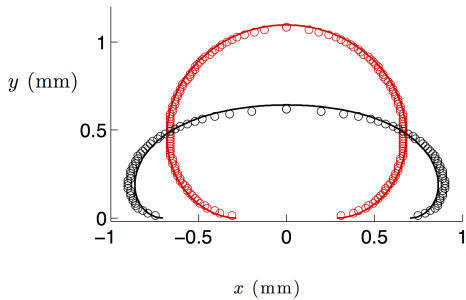}
	\caption{\footnotesize Shape of an oxygen drop of volume $\Omega=1.2\;$mm$^3$ in the absence (gray - red online) and presence (black) of a gradient of squared magnetic field of $70\;$T$^2/$m (obtained at a distance $z=1\;$mm above a magnet). Solid lines are solutions of equation \eqref{eq:courbure} (derived in the appendix) and data (circles) are obtained from Figures~\ref{fig:StaticDeformation}a and~\ref{fig:StaticDeformation}b.}
	\label{fig:CalculForme}
\end{figure}

The exact shape of a drop in the presence of a vertically varying magnetic field is determined by a balance between hydrostatic, magnetic and capillary pressures. This yields a differential equation for the drop profile, as detailed in the appendix. For the magnets that we used, the field decreases on a centimetric length-scale, which is much larger than the millimetric drop thickness. Therefore, the gradient of magnetic force is negligible at the scale of the drop, and shapes are those of puddles in a uniform enhanced gravitational field $g + f_m/\rho$ (Fig.~\ref{fig:CalculForme}). However, in the presence of a very large gradient of magnetic field, we could expect different shapes, with an enhanced curvature at the bottom. As discussed in the appendix, such shapes present similarities with what is observed in electrowetting on highly hydrophobic surfaces \cite{Mugele2007}. Note that we do not discuss here the shape of the interface below the drop, that may be deformed by the pressure in the vapour layer \cite{Nagel2012}. The vapour layer itself should be affected by the magnetic field, but the effect should be negligible: as recalled in \cite{Quéré2013}, its thickness $\delta$ scales as $(R b)^{1/2}$ where the length $b$ depends on the applied field, as $f_m^{-1/4}$ for $f_{m} \gg \rho g$. But the field also increases the drop radius, as $f_m^{1/4}$ in the same limit. Hence the film thickness is independent of $f_m$, when the field is strong, which qualitatively explains why we never saw any collapse of the Leidenfrost state, even in the limit $f_m \gg \rho g$.

\section{Capturing drops}

\begin{figure}
\centering

\includegraphics[width=15cm]{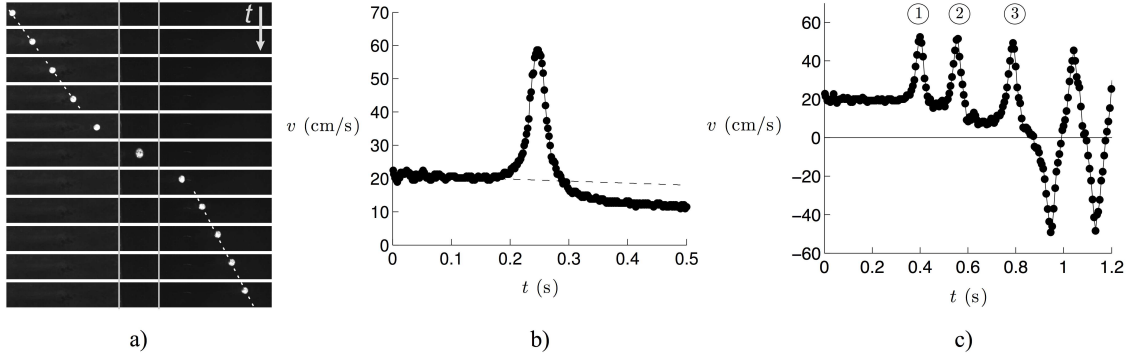}
	\caption{\footnotesize a) Top views of an oxygen drop of radius $R=1\;$mm passing above a parallelepipedic magnet of width $1\;$cm indicated by the vertical grey lines. Time increases from top to bottom ($\Delta t=25\;$ms between two images). The drop is deformed above the magnet and it comes out of the magnetic trap significantly slowed down, as seen from the change of slope of the dotted lines. (enhanced online, movie is slowed down 20 times). b) Velocity of the drop as a function of time. The drop arrives at $V=20\;$cm/s, and it is successively accelerated and decelerated above the magnet, which it leaves at a velocity  $V'=12\;$cm/s. c) Measurement of the velocity of a drop passing above a series of three magnets (indicated by their number). The drop loses some speed at each magnet and it is finally captured above the third magnet: the velocity goes to zero and the drop then oscillates in the trap. (enhanced online, movie is slowed down 5 times).}
	\label{fig:ManipDynamique}
\end{figure}

Since there is no contact between liquid oxygen and its support, friction in the Leidenfrost state is almost inexistant: several meters are needed to observe the deceleration of a millimetric Leidenfrost drop thrown on a horizontal surface at a few tens of cm/s \cite{Dupeux2011}. Oxygen drops being sensitive to magnetic fields, it is natural to wonder whether and how magnetic traps can affect their mobility. To answer these questions, the following experiment is conducted: an oxygen drop of typical radius $R=1\;$mm is thrown tangentially at a velocity $V$ on a horizontal square glass plate of side $10~\mathrm{cm}$ and thickness $2~\mathrm{mm}$, under which is placed a parallelepipedic neodymium magnet (square cross-section of $1\;\mathrm{cm^2}$ and length of $4\;$cm), perpendicularly to the trajectory of the drop. Figure \ref{fig:ManipDynamique}a is a spatio-temporal picture of such an experiment, seen from above, for which the drop arrives in the magnetic trap at a velocity $V=20\;$cm/s. 

We show in Figure \ref{fig:ManipDynamique}b the oxygen velocity $v$ as a function of time during the experiment. It first weakly decreases far from the magnet. The corresponding deceleration can be extracted from the data slope between $t=0\;$s and $t=0.15\;$s (dashed line in Figure \ref{fig:ManipDynamique}b). It is equal to $6\;$cm/s$^2$, so that the friction is of order  $\rho R^3 \,\mathrm{d} V / \mathrm{d}t \sim 0.1\;\mu$N. Comparatively, the inertial friction in air is $\rho_{air} V^2 R^2$, on the order of 0.01 $\mu$N to 0.1 $\mu$N for $V$ between 20~cm/s and 60~cm/s. One also has to consider the viscous friction in the vapour layer, which is  $\eta_{air} V R^2 / \delta \sim 0.05~\mu$N (for a layer of thickness $\delta \sim 50~\mu$m, evaluated from close-up photographs taken with a stereo microscope). All together, this gives a friction force consistent with our measurement. As the drop gets closer to the magnet, it first accelerates to reach a velocity of about $V_{max} \approx 60\;$cm/s. The reduction of magnetic energy $|E_m| = \chi B^2 / 2 \mu_o$  (eq.~\ref{eq:Em}), which typically is $350~\mathrm{J/m}^3$ for a drop in a field of magnitude $B=0.5~T$, is on the order of the increase of kinetic energy per unit volume $\rho (V_{max}^2 - V^2)/2 \simeq 180~\mathrm{J/m}^3$, yet higher since some magnetic energy is also transferred into deformation of the drop. The drop then decelerates as it leaves the magnetic trap, from which it comes out with a velocity $V'=12\;$cm/s, significantly below the velocity expected in the absence of a magnet. Passing across the trap thus produces on the whole a significant loss of kinetic energy, by a factor of order 3 in this example.

\begin{figure}
	\centering
		   \includegraphics[height=5cm]{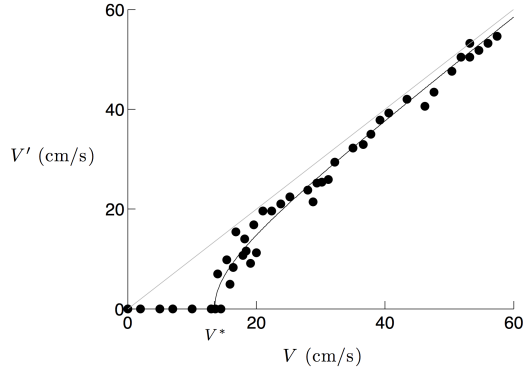}
	\caption{\footnotesize Exit velocity $V'$ of an oxygen drop of radius $R=1\;$mm escaping a magnetic trap, as a function of its initial velocity $V$. We observe a critical velocity $V^*=13.5\;$cm/s below which the drop is captured by the magnet ($V'=0$). The solid line represents equation \ref{eq:Vstar}, and the grey thin line is $V'=V$.}
	\label{fig:vstar}
\end{figure}

The same experiment can be done using a series of parallel magnets (fig.~\ref{fig:ManipDynamique}c). A certain amount of kinetic energy is lost above each magnet, and the drop gets captured when its inertia becomes weaker than the magnetic attraction. The drop is then trapped ($<v(t)>\,=0$) and it oscillates several times above the magnet before viscosity and friction of air damp the motion. This experiment is repeated for various initial velocities between 1 and $60\;$cm/s. Figure \ref{fig:vstar} summarizes the results, representing for each magnetic trap the exit velocity $V'$ as a function of the entrance velocity $V$. There is a critical velocity $V^*=13.5\pm1.0\;$cm/s below which the drop is not able to escape ($V'=0$). Above this value, the difference $\Delta V=V-V'$ decreases as $V$ increases, so that $\Delta V/V$ tends to zero for $V\gg V^*$.

\begin{figure}
	\centering
		   \includegraphics[width=10cm]{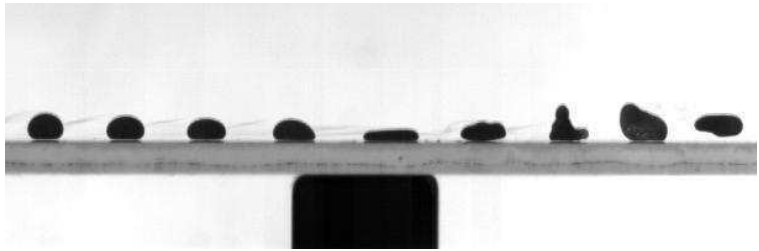}
	\caption{\footnotesize Side view chronophotograph of an oxygen drop of radius $R=1\;$mm passing above the magnet (black rectangle below the glass plate) at a velocity $V=60\;$cm/s. Time interval between successive photos: $8\;$ms. (enhanced online, movie is slowed down 50 times).}
	\label{fig:ExpSide}
\end{figure}

As it crosses the magnet (of centimeter-size), a drop loses a negligible amount of momentum ($\Delta V / V \simeq 1 \%$) because of friction due to the surrounding air. The special dissipation in the trap rather seems to originate from the deformations in the field, which are not converted back into horizontal velocity when the drop leaves the magnet but relax into vibration modes \cite{Brunet2011}, which are eventually damped. This is obvious when looking at experiments from the side, as shown in Figure \ref{fig:ExpSide}: the drop is flattened above the magnet and then it retracts to recover its rounded shape, which induces large vibrations and even lead in that case to a small jump, showing that the energy transferred into deformation can, at best, be converted into vertical kinetic energy. 

We assume that the energy $E_d$ stored into deformation is lost, which yields a critical velocity for which the drop cannot escape $V^*=\sqrt{2 E_d/m}$, where $m\ =4 \pi \rho R^3 /3$ is the mass of the drop. To evaluate $E_d$, we look at the maximal radius $R_{max}$ achieved by the drop while crossing a magnet. We measure a ratio $R_{max}/R=1.3$, which does not seem to depend on the drop velocity for $V$ between $1$ and $60\;$cm/s. This value is similar to the one expected in a static situation: the numerical resolution of the shape (eq.~\ref{eq:courbure}) at $z=2\;$mm above the same magnet gives $R_{max}/R\simeq 1.26$. The deformation is thus equivalent when the drop crosses the magnet and when it is at rest in a similar magnetic field. Of course, if the time needed to deform the drop $\tau_1\sim\sqrt{\rho R^3/\gamma}$ becomes larger than the time needed to cross the magnet $\tau_2\sim L/V$ (where $L$ is the size of the magnet), the drop does not have time to reach its maximal deformation. This would be the case for drops with a velocity $V > \sqrt{\gamma L^2/\rho R^3}\sim 1$ m/s, a value that is never reached in our experiments.

The surface energy stored in deformation is $E_d\approx\gamma\delta \Sigma$, where $\delta \Sigma$ is the increase in surface area of the drop. For the sake of simplicity, we restrict the discussion to spherical droplets, for which the variation of surface area is of order $\delta \Sigma \simeq 4 \pi (R+\delta R)^2 - 4 \pi R^2 \simeq 4\pi \left(2R\delta R + \delta R ^2\right)$, where $\delta R = R_{max}-R$. For small deformations ($\delta R < R$), we obtain $E_d \simeq 8\pi\gamma R\delta R$, which gives a critical velocity:
\begin{equation}
 V^* \simeq \sqrt{\frac{12\gamma \delta R}{\rho R^2}} = 20\;\mathrm{cm/s}
 \label{eq:Vs}
 \end{equation}
 \noindent
 on the order of the measured value of $13.5~\mathrm{cm/s}$. Above this threshold velocity, the drop is able to escape but loses a certain amount of kinetic energy which is always the same since $\delta R$ does not depend on the velocity, in the range we explored. The terminal velocity $V'$ can then be written:
 
 \begin{equation}
 V' = \sqrt{V^2 - V^{*2}}
 \label{eq:Vstar}
 \end{equation}
 \noindent
 This equation is plotted in Figure \ref{fig:vstar} with $V^*=13.5\;\mathrm{cm/s}$ and it matches the measurements. It predicts in particular a critical behavior near $V=V^*$ and the asymptotic behavior ($V'\approx V$) in the limit $V \gg V^*$ .

\begin{figure}
	\centering
		   \includegraphics[height=5cm]{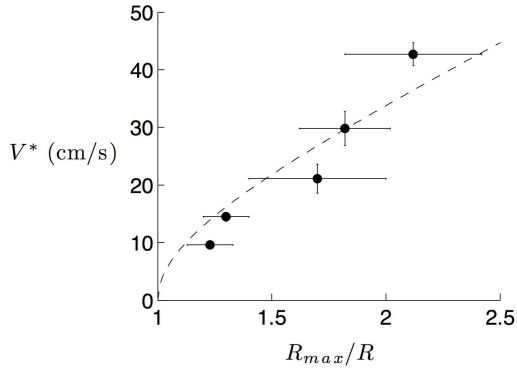}
	\caption{\footnotesize Critical velocity $V^*$ as a function of the deformation $R_{max}/R$. The dashed line represents equation \ref{eq:vstarfull} with  $\left( 6 \gamma/\rho R \right)^{1/2}=19.5\;\mathrm{cm/s}$, adjusted to fit the data.}
	\label{fig:VstarRmax}
\end{figure}

 One way to test the scaling law \eqref{eq:Vs} for $V^*$ is to vary $\delta R$ by changing the distance between the magnet and the glass plate. We used five different distances $z$, between $0.3\;\mathrm{mm}$ and $3\;\mathrm{mm}$, and measured for each of them $R_{max}$ and $V^*$. Deformation as high as $R_{max}/R \approx 2$ were obtained, for which the critical velocity $V^*$ is approximately four times higher than before, as shown in Figure \ref{fig:VstarRmax}. Since $\delta R \sim R$, the approximation of small deformation is not valid anymore. In that regime, the general equation for the critical velocity $V^* = \sqrt{2 E_d /m}$ becomes:
 
 \begin{equation}
 V^* \sim \sqrt{\frac{6 \gamma}{\rho R} \; \left(  2 \left(\frac{R_{max}}{R}  -1 \right) + \left(\frac{R_{max}}{R}  -1 \right)^2 \right)  }
 \label{eq:vstarfull}
 \end{equation}
  The dotted line in Figure \ref{fig:VstarRmax} represents this equation where the velocity $\sqrt{ 6 \gamma/\rho R }$ is treated as an adjustable parameter. The best fit is obtained for a value of $19.5\;\mathrm{cm/s}$, consistent with the expected order of magnitude, that is, $26\;\mathrm{cm/s}$. 
 
\begin{figure}[h!]
	\centering
		   \includegraphics[width=12cm]{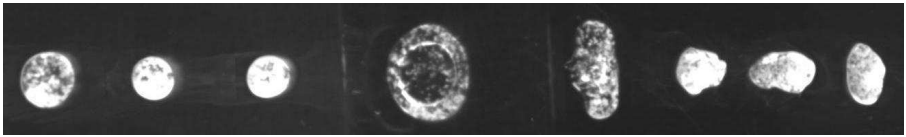}
	\caption{\footnotesize Top view chronophotograph of an oxygen drop of radius $R=1\;$mm passing above a magnet located at $z=0.5\;$mm below the liquid, and indicated by the white lines. Initial velocity of the drop is $V=48\;$cm/s. The variation of $R$ in the three first pictures is due to small initial vibration of the drop. Interval between images: $10\;\mathrm{ms}$. (enhanced online, movie is slowed down 60 times).}
	\label{fig:ChronoDef}
\end{figure}

Imposing a large deformation to the drop leads to a significant loss of kinetic energy. At an altitude $z=2$ mm above the magnet, the deformation is small and similar to the one observed in the static case. This is not true anymore for large deformations, as can be seen in Figure \ref{fig:ChronoDef}, showing a drop passing at $z=0.5\;\mathrm{mm}$ above the magnet. We observe the formation of a rim, comparable to the shape of a bouncing drop \cite{Roisman2009}, which is clearly not expected in a static situation. 

\noindent
We measure a deformation $\delta R \sim R = 1~\mathrm{mm}$ much larger than the static value of $\delta R\simeq 0.3$ mm given by the resolution of equation \eqref{eq:courbure} in the same field. The expansion takes place in approximately $\delta t \sim 10\;\mathrm{ms}$, which gives a characteristic internal velocity of $10\;\mathrm{cm/s}$. The energy  of this flow is thus of order $\rho (R^2/\delta t^2) R^3$, that we can compare to the surface energy $\gamma \delta \Sigma \sim \gamma R^2$. Their ratio gives a Weber number of order unity for $\rho=10^3\;\mathrm{kg/m^3}$, $R=1\;\mathrm{mm}$ and $\gamma\sim10\;\mathrm{mN/m}$:

\begin{equation}
\mathrm{We}\sim \frac{\rho R^3}{\gamma \delta t^2} \sim 1
\end{equation}

The inertia of the fluid is thus on the order of surface forces which explains why the radius of the drop can become larger than in the static case. In the regime of small deformation, mainly studied here, this number is one order of magnitude smaller, so that equations \eqref{eq:Vs} and \eqref{eq:Vstar} can be applied.

 \section{Conclusion}
We described how drops of liquid oxygen can be deformed using a magnetic field. We characterized the magnetic force and showed that it can be several times higher than gravity and hence flatten the drop. This effect has been accounted for by the modification of the capillary length due to the magnetic force. We determined the equation for the shape of a drop in the presence of a vertically varying magnetic field and verified its validity by comparing the results to the observed shapes.
 
We also pointed out that these deformations can explain the strong deceleration of oxygen drops passing above a magnet. The surface energy stored in deformation is not recovered into kinetic energy, but rather converted into surface vibrations and internal motion, that are eventually damped. Hence, although the magnetic force is conservative, it can induce dissipation because of the deformability of the drop. We show that there is a critical velocity below which the drop is not able to escape from the magnetic trap. We proposed a scaling law that accounts for it. The magnets we used can rapidly impose large deformations to the liquid: we observed cases where a rim appears during the deformation, such as those observed with drops impacting a solid surface.

This simple system allows one to probe the dynamics of drops in a controlled way. We can imagine other experiments in the same vein using liquid oxygen or other magnetic fluids such as aqueous ferrofluids in the Leidenfrost state: imposing an oscillating magnetic field could be used for example to study the vibration modes of a droplet. A revolving field would permit the rotation (and subsequent deformation) of a liquid. Finally, this kind of control might be of high interest in application to aerospace engineering (where liquid oxygen is used as a propellant in combination with liquid hydrogen or kerosene), to control the position of liquid oxygen near the outlet of a rocket combustion chamber.

\bigskip
\noindent

\begin{appendix}
\section*{Appendix : calculation of the shape of a drop under magnetic field}
\renewcommand{\theequation}{A\arabic{equation}}  
\setcounter{equation}{0}  
\renewcommand{\thefigure}{A\arabic{figure}} 
\setcounter{figure}{0}  

\begin{figure}[!h]
	\centering
		   \includegraphics[height=3.5cm]{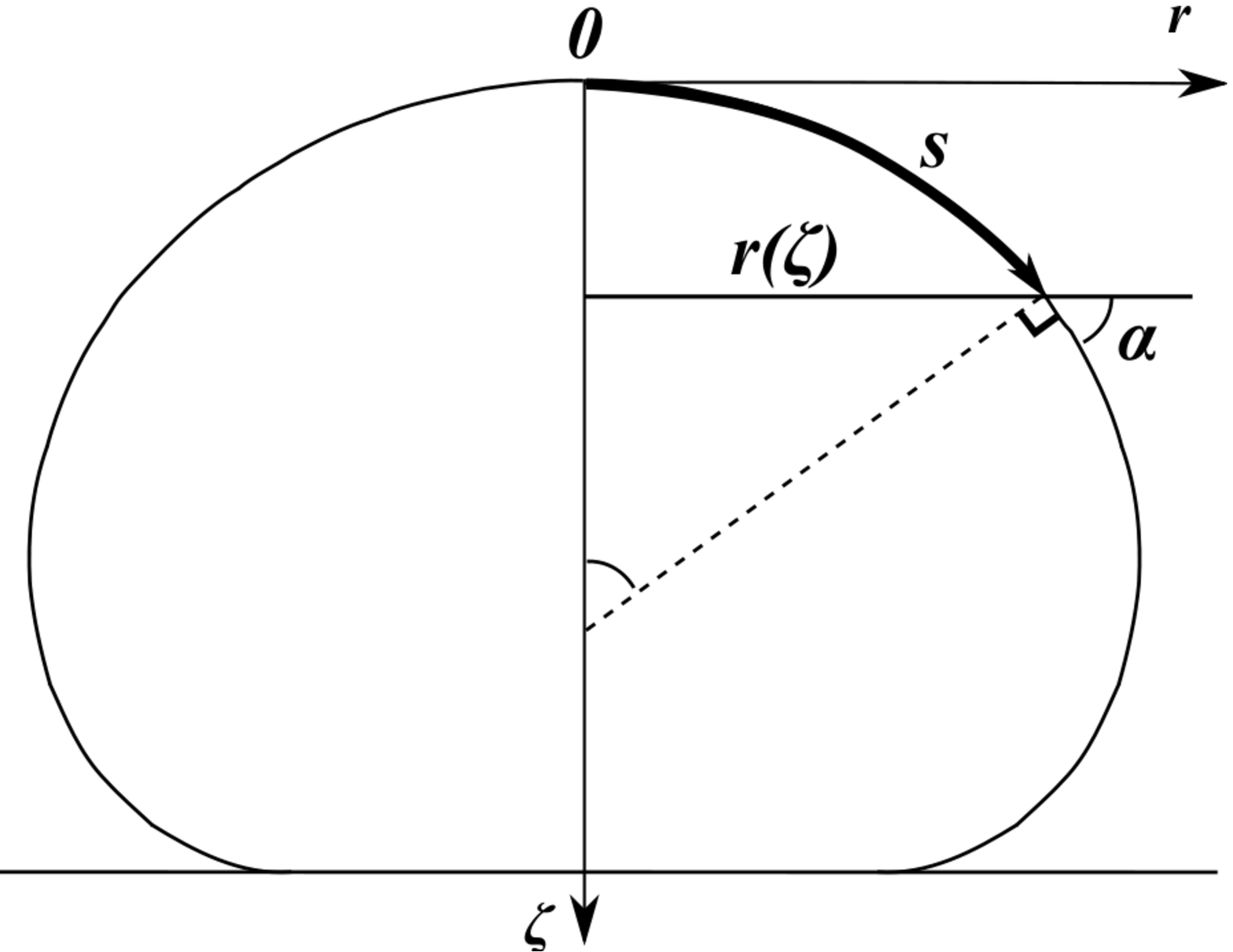}
	\caption{\footnotesize Diagram of the drop. $s$ is the curvilinear coordinate along the profile of equation $r(\zeta)$ defining the shape of the drop.}
	\label{fig:SchemaForme}
\end{figure}

\begin{figure}
 		\centering
  		\includegraphics[height=6cm]{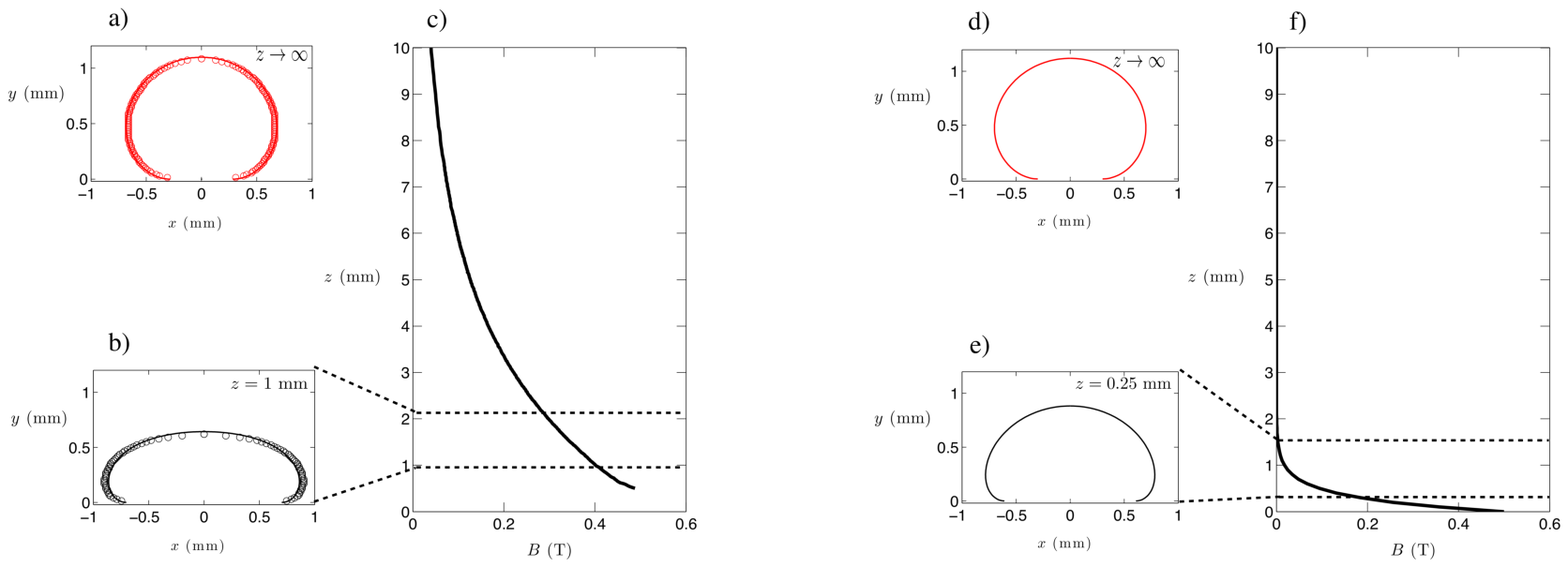}
	\caption{\footnotesize a) Shape of an oxygen drop of volume $\Omega=1.2\;$mm$^3$ in the absence of magnetic field. Solid line is the solution of equation \eqref{eq:courbure} and circles are obtained from Figure ~\ref{fig:StaticDeformation}a. b) The same drop placed at $z=1$ mm above the magnet. Similarly, solid line comes from computation and circles are obtained from Figure ~\ref{fig:StaticDeformation}b. c) Magnetic field $B$ as a function of height $z$ above a centimetric cylindrical magnet. We do not have data below $z=0.5$ mm, which corresponds to half the thickness of the probe used for the measurement. d) Computed shape of a drop of radius $\Omega=1.2\;$mm$^3$ in the absence of magnetic field. e) Computed shape of the same drop placed in a hypothetical magnetic field with a very strong gradient. The magnetic force is much higher at the bottom of the drop that at the top, yielding a shape with an enhanced curvature at the bottom, which is very different from the one presented in b). e) Hypothetical magnetic field with a very strong gradient. The corresponding equation is $B(z)=0.5\,\exp(-z/0.3)$ (with $z$ expressed in mm).}

	\label{fig:champ_forme}
\end{figure}

The shape of a drop of liquid oxygen in a gradient of magnetic field can be calculated using the pressure balance on the drop. In order to simplify calculations, we place the origin at the upper pole of the drop, which is the starting point of the calculation, as shown in Figure \ref{fig:SchemaForme}. The vertical axis is oriented downwards. We consider the drop as axisymmetric and defined by a profile $r(\zeta)$ giving the radius of the drop at a given height $\zeta$. 
\noindent
We call $s$ the curvilinear coordinate along $r(\zeta)$ and $\alpha$ the angle between the surface and the horizontal. The pressure at a height $\zeta$ is:
\begin{equation}
P(\zeta)=P(0)+\rho g \zeta + \int_0^{\zeta} f_m(\zeta') \mathrm{d}\zeta'
\label{eq:hydrostat}
\end{equation}
\noindent
The drop shape is given by balancing this pressure with the capillary pressure:
\begin{equation}
P(\zeta)=P_{atm}+\gamma \kappa(\zeta)
\label{eq:Laplace}
\end{equation}
\noindent
where $\kappa(\zeta)$ is the mean curvature of the surface at a height $\zeta$. For $\zeta=0$, we obtain $P_{atm}+\gamma \kappa_o=P(0)$, where $\kappa_o$ is the curvature at the top of the drop. Combining equations \eqref{eq:hydrostat} and \eqref{eq:Laplace} with the fact that $f_m$ derives from the magnetic energy $E_m$ (eq.~\ref{eq:Em}), we obtain:
\begin{equation}
\kappa(\zeta)=\frac{\zeta}{a^2} + \kappa_o + \frac{\chi}{2 \mu_o \gamma}\left[B^2(\zeta)-B^2(0)\right]
\label{eq:courbure}
\end{equation}
In order to compute the drop shape from this equation, we use the relationship $\kappa(\zeta)=\mathrm{d} \alpha /\mathrm{d}s + \sin \alpha/r$, combined with $\mathrm{d} r /\mathrm{d}s = \cos \alpha$ and $\mathrm{d} \zeta /\mathrm{d}s = \sin \alpha$. Hence, we obtain a system of three differential equations for $\alpha$, $r$ and $\zeta$ that we can solve numerically with appropriate boundary conditions (in particular, a contact angle of 180$^{\circ}$ at the solid contact) and compare to the measured profiles. The comparison is presented in Figure \ref{fig:champ_forme}. The grey curve (red online, fig.~\ref{fig:champ_forme}a) represents the shape of a drop of radius $R=0.7$ mm in the absence of magnetic field ($z\rightarrow \infty$). Circles are measurements made from Figure \ref{fig:StaticDeformation}a. We fit this shape using equation~\eqref{eq:courbure}, which fixes the value of $\kappa_o$. We then calculate the shape of a drop of same volume placed at an altitude $z=1$ mm above a magnet (fig.~\ref{fig:champ_forme}b). The value of $B(\zeta)$ used in the calculation comes from the measurement performed with the teslameter (fig.~\ref{fig:champ_forme}c).

In the presence of a very large gradient of magnetic field, we could expect the magnetic force to be higher at the bottom of the drop than at the top, inducing an enhanced curvature at the bottom, as shown in Figure \ref{fig:champ_forme}e. To compute this shape, we used a hypothetical field $B(z)=0.5\,\exp(-z/0.3)$ (where $z$ is expressed in mm and $B$ in T), which presents a strong variation on the drop scale (fig.~\ref{fig:champ_forme}f). Such a shape is clearly different from the puddles obtained under a uniform magnetic force, and it has similarities with shapes observed in electrowetting on highly hydrophobic surfaces, for which the connection between the local contact angle fixed by the Young--Laplace equation and the curvature imposed by the electric field similarly generates a strong curvature at the bottom \cite{Mugele2007}.

\end{appendix}

\bibliographystyle{unsrt}

\begin{thebibliography}{99}

\bibitem{Dewar1927}{J. Dewar, \emph{Collected papers of Sir James Dewar} (Cambridge University Press, Cambridge, 1927).}

 \bibitem{Leidenfrost1756}{J. G. Leidenfrost \emph{De Aquae Communis Nonnullis Qualitatibus Tractatus} (Duisburg, 1756).}
 
 \bibitem{Gottfried1966}{B. S. Gottfried, C. J. Lee and K. J. Bell, ``The Leidenfrost phenomenon: film boiling of liquid droplets on a flat plate,'' Int. J. Heat Mass Transfer. \textbf{9}, 1167 (1966).}
  
 \bibitem{Thoroddsen2012}{I. U. Vakarelski, N. A. Patankar, J. O. Marston, D. Y. C. Chan and S. T. Thoroddsen, ``Stabilization of Leidenfrost vapour layer by textured superhydrophobic surfaces,'' Nature, \textbf{489}, 274 (2012).}
 
 \bibitem{Quéré2013}{D. Qu\'er\'e, ``Leidenfrost Dynamics,'' Annu. Rev. Fluid. Mech., \textbf{45}, 197 (2013).}
 
 \bibitem{Catherall2003}{A. T. Catherall, L. Eaves, P. J. King and S. R. Booth, ``Floating gold in cryogenic oxygen,'' Nature \textbf{422}, 579 (2003).}
 
 \bibitem{Nikolayev2009}{G. Pichavant, B. Cariteau, D. Chatain, V. Nikolayev, and D. Beysens, ``Magnetic compensation of gravity: experiments with oxygen,'' Microgravity Sci. Tech. \textbf{21}, 129 (2009).}
 
 \bibitem{Takeda1991}{ M. Takeda and K. Nishigaki, ``Shape deformation of the gas-liquid interface of liquid oxygen in high-magnetic fields,'' Phys. Rev. A \textbf{43}, 2081 (1991).}

 \bibitem{Catherall2003Instabilities}{ A. T. Catherall, K. A. Benedict, P. J. King and L. Eaves, ``Surface instabilities on liquid oxygen in an inhomogeneous magnetic field,'' Phys. Rev. E \textbf{68}, 037302 (2003).}

\bibitem{Youngquist2003}{R. C. Youngquist, ``Dynamics of a finite liquid oxygen column in a pulsed magnetic field,'' IEEE transactions on magnetics \textbf{39}, 2068 (2003).}

\bibitem{Boulware20102}{J. C. Boulware, H. Ban, S. Jensen and S. Wassom, ``Experimental studies of the pressures generated by a liquid oxygen slug in a magnetic field,'' J. Mag. Mag. Mat. \textbf{322}, 1752 (2010).}

\bibitem{Takeda1991ST}{M. Takeda and K. Nishigaki, ``Measurement of the surface tension of liquid oxygen in high magnetic fields,'' J. Phys. Soc. Japan \textbf{61}, 3631 (1991).}

\bibitem{Maha1999}{L. Mahadevan and Y. Pomeau, ``Rolling droplets,'' Phys. Fluids \textbf{11}, 2449 (1999).}

\bibitem{Rosensweig}{R. E. Rowensweig \emph{Ferrohydrodynamics} (Dover publications, Mineola, 1985).}
\bibitem{Mugele2007}{F. Mugele and J. Buehrle, ``Equilibrium drop surface profiles in electric fields,'' J. Phys.: Condens. Matter  \textbf{19}, 375112 (2007).}
\bibitem{Nagel2012}{J. C. Burton, A. L. Sharpe, R. C. A. van der Veen, A. Franco and S. R. Nagel, ``Geometry of the vapor layer under a Leidenfrost drop,'' Phys. Rev. Lett.  \textbf{109}, 074301 (2012).}
\bibitem{Dupeux2011}{G. Dupeux, M. Le Merrer, C. Clanet and D. Qu\'er\'e, ``Trapping Leidenfrost drops with crenelations,'' Phys. Rev. Lett. \textbf{107}, 114503 (2011).}

 \bibitem{Brunet2011}{P. Brunet and J. H. Snoeijer, ``Star-drops formed by periodic excitation and on an air cushion - A short review'', E. Phys. J. Special Topics, \textbf{192}, 207 (2011).}


\bibitem{Roisman2009}{I. V. Roisman, E. Berberovic and C. Tropea, ``Inertia dominated drop collision,'' Phys. Fluids \textbf{21}, 052103 (2009).}


\end{thebibliography}

\end{document}